\documentclass[12pt]{article}
\usepackage{a4wide}
\usepackage{hyperref}
\usepackage{amssymb,amsmath}
\usepackage[dvips]{graphicx,color}
\begin{document}
{\renewcommand{\thefootnote}{\fnsymbol{footnote}}

\begin{center}
{\LARGE On the geometry of no-boundary instantons in \\ loop quantum cosmology}\\
\vspace{1.5em}
Suddhasattwa Brahma$^1$\footnote{e-mail address: {\tt
    suddhasattwa.brahma@gmail.com}}  
and Dong-han Yeom$^{1,2}$\footnote{e-mail address: {\tt innocent.yeom@gmail.com}}
\\
\vspace{0.5em}
$^1$Asia Pacific Center for Theoretical Physics, Pohang 37673, Korea\\
$^2$Department of Physics, POSTECH, Pohang 37673, Korea\\
\vspace{1.5em}
\end{center}
}

\renewcommand{\d}{\text{d}}
\setcounter{footnote}{0}

\begin{abstract}
We study the geometry of Euclidean instantons in loop quantum cosmology (LQC) such as those relevant for the no-boundary proposal. Confining ourselves to the simplest case of a cosmological constant in minisuperspace cosmologies, we analyze solutions of the semiclassical (Euclidean) path integral in LQC. We find that the geometry of LQC instantons have the peculiar feature of an infinite tail which distinguishes them from Einstein gravity. Moreover, due to quantum-geometry corrections, the small-$a$ behaviour of these instantons seem to naturally favor a closing-off of the geometry in a regular fashion as was originally proposed for the no-boundary wavefunction.
\end{abstract}

\section{Introduction}
Recently, the introduction of the `no-boundary' proposal in loop quantum cosmology (LQC), for minisuperspace models, has unveiled a lot of interesting physical possibilities \cite{Brahma:2018elv}. It has been shown that the original Hartle-Hawking formulation \cite{Hartle:1983ai}, improved by an effective action which includes corrections due to LQC, can lead to an expanded solution space due to singularity-resolution \cite{Bojowald:2006da} coming from the latter. In particular, it has been shown that not only is the probability for a de-Sitter (dS) universe nucleating from nothing increased in such a scenario, there can now be compact, non-singular instantonic solutions in cases where there were none in Einstein gravity. As an example, the model of a Friedmann-Robertson-Lemaitre-Walker (FLRW) closed universe, coupled to a massless scalar field,  was considered in \cite{Brahma:2018elv} and shown to have a nontrivial compact instanonic solution with a finite probability for nucleation. This study has opened the doors for revisiting the original no-boundary proposal augmented by quantum-geometry effects governing the dynamics of the early-universe which are, in any case, expected for a meaningful UV-completion. A detailed study of such effects for physically relevant questions such as the probability of inflation and number of $e-$folds predicted by the (improved) no-boundary measure can now be answered within the purview of LQC. However, this is not the intention of present work and shall be pursued later elsewhere.

In this work, we focus on the geometry of these  Euclidean instanton solutions in LQC. This is a necessary first step before using such solutions to consider nucleation of universes from nothing and employing the measure provided by the associated wavefunction for predicting the probabilities of physically interesting phenomena for the Lorentzian histories. Our starting point shall be the (Euclidean) path integral for quantum gravity, along with the prescription that the initial conditions are provided by the no-boundary proposal. The main new ingredient, in comparison to the original Hartle-Hawking proposal, shall be the `effective' action appearing in the path integral derived from LQC, as opposed to the usual Einstein-Hilbert one. (It was established in \cite{Ashtekar:2010gz} that by replacing the standard FLRW action by the `polymerized' version of it, the path integral formulation of LQC, in its phase space realization, retains all the crucial aspects of the quantum geometry which appear in the canonical LQC.) Other than this, our formalism shall be exactly the same as in the original no-boundary proposal: We shall look only at the saddle-point approximation of the Euclidean path integral and consider the wavefunction to be a functional of the value of the scale factor only at the final (spatial) boundary. Moreover, as shall be obvious throughout our paper, we make a minisuperspace approximation for all our calculations and consider the matter content to be only that due to a cosmological constant. The latter approximation ensures that we always have a compact instantonic solution and do not require dealing with subtleties which can give rise to Euclidean wormholes \cite{Chen:2016ask}. Since we want to show new features of the LQC instantons with respect to its geometry, as compared to the original Hartle-Hawking ones, these approximations shall help us emphasize our main result without unnecessarily complicating the system.

For our purposes, the effective action consists of two main types of quantum corrections specific to LQC -- the holonomy and inverse-triad modifications \cite{Bojowald:2002gz}. The first appears due to the fact that there are no quantum operators corresponding to the connection or extrinsic curvature (in other words, the momenta conjugate to the spatial metric) on the kinematic\footnote{`Kinematic' here refers to the fact that the Gauss and the (spatial) diffeomorphism constraints have been solved whereas `physical' would imply the solution of the Hamiltonian constraint as well. For a minisuperspace model, these distinctions are not very important since the only leftover symmetry of the system is time-reparameterization invariance.} Hilbert space of the theory. On the other hand, there are well-defined operators corresponding to the holonomy (or parallel transport) of the connection \cite{Ashtekar:1995zh}. Therefore, one expresses the curvature operator in terms of these holonomies instead of the connection itself. Classically, one can take the limit such that one recovers the expression of the curvature written in terms of the  connection from the expression given for the holonomies. However, the geometrical operators in the full loop quantun gravity have discrete spectra for quantities such as area and volume \cite{Ashtekar:1996eg} on the kinematical Hilbert space spanned by the spin-network states, rendering taking such a limit unviable. Therefore, one inherits an `area-gap', in analogy with the minimum energy-gap of the harmonic oscillator, from the full theory in LQC \cite{Ashtekar:2003hd}. The main effect of this regularization of the curvature in terms of holonomies, for symmetry-reduced models, lie in replacing the extrinsic curvature by matrix elements of $SU(2)$-holonomies which are periodic functions of the connection. Specifically, for minisuperspace cosmologies, we have $H \rightarrow \sin(\delta H)/\delta$, where $H$ is the Hubble parameter and $\delta$ is related to the area-gap.

The discrete spectra of area and volume operators also lead to other type of corrections in LQC. The most significant of them are the inverse-triad corrections which arise from the requirement of having a well-defined operator corresponding to the inverse of some power of the scale factor whose spectra contains the zero eigenvalue. Naively, it is impossible to have a densely-defined operator in such a case. However, using the aforementioned holonomy operators and what is commonly known as the `Thiemann trick' in the literature, one can express the relation \cite{Thiemann:1997rt}
\begin{eqnarray}
	\hat{h}^{-1}[\hat{h},\sqrt{\hat{a}}]= -\frac{1}{2}\hbar \delta	\widehat{a^{-1/2}}\,.
\end{eqnarray}
In this definition,  $\hat{h} := \widehat{\exp(i\delta p_a)}$, for the momentum conjugate to the scale factor $p_a\propto \dot{a}$, is precisely a $SU(2)$-valued holonomy operator mentioned previously. It is clear from this relation that one can have an operator, whose classical limit is some inverse power of the scale factor on the RHS although we do not require any inverse operator on the LHS \cite{Bojowald:2002ny}. Using this, one gets rid of the singular behaviour of any function which contains some inverse power of $a$ due to the replacement by these aforementioned inverse-triad corrections. Once again, their form for minisuperspace cosmologies is rather simple, as shall be explicitly demonstrated later.

Let us briefly summarize our main result. Conceptually, at least in the cosmological constant case, the main effect of the LQC quantum-geometry corrections lies in the small-$a$ behaviour of the Euclidean LQC instantons. As shall be demonstrated, due to the inverse-triad corrections, the LQC-modified Friedmann equation is such that the solution tails off to zero at the symmetry point of the theory. The geometry of the LQC instanton will emerge to be quite different from the original Hartle-Hawking proposal with an infinitely stretched tail in Euclidean time; however, their \textit{topology} remains the same. Moreover, such an infinitely long tail of the instanton (in imaginary time) is not an inherent problem since the only meaningful physical quantity is the  probability  of nucleation which remains finite for this system. The interesting fact is the quantum-geometry regularization is such that this tail closes the geometry in a regular way without requiring any additional fine-tuning even though the field equations are heavily modified in LQC. This is suggestive of the fact that the no-boundary proposal is robust and, if anything, such a necessary tail-off of LQC instantons to zero points towards it being more natural in the presence of quantum-geometry corrections.

\section{The Hartle-Hawking proposal revisited}
In this section, we first briefly review the geometry of the no-boundary instantons in Einstein gravity. In the process, we also fix our notation for the rest of the paper.

\subsection{The Wheeler-de Witt equation and boundary conditions}
Let us consider Einstein gravity with a scalar field
\begin{eqnarray}
S = \int \sqrt{-g}\, \d x^{4} \left[ \frac{\mathcal{R}}{16\pi} - \frac{1}{2} \left(\nabla \phi\right)^{2} - V(\phi) \right]\,.
\end{eqnarray}
In this paper,  we shall exclusively focus on a minisuperspace cosmological model \cite{Vilenkin:1987kf}
\begin{eqnarray}
\d s^{2} = \sigma^{2} \left[ - N^{2}(t) \d t^{2} + a^{2}(t) \d\Omega_{3}^{2} \right]\,,
\end{eqnarray}
where $\sigma^{2} = 2/3\pi$ is some normalization constant\footnote{Note that we choose this normalization at this point for historical reasons and to keep the resulting equations simple. However, we shall change this normalization later on to facilitate comparison with LQC.}. By assuming the slow-roll limit $\dot{\phi} \approx 0$, the Lagrangian can be simplified as
\begin{eqnarray}
\mathcal{L} = \frac{1}{2} N \left[ a \left( 1 - \frac{\dot{a}^{2}}{N^{2}} \right) - \bar{V} a^{3} \right],
\end{eqnarray}
where $\bar{V} = 16 V / 9$. From this, one can get the conjugate momentum $p_{a} = - a \dot{a} / N$. The Hamiltonian $\mathcal{H}$ is obtained by the usual Legendre transform
\begin{eqnarray}
\mathcal{L} = p_{a} \dot{a} - N \mathcal{H}\,,
\end{eqnarray}
where
\begin{eqnarray}
\mathcal{H} = - \frac{1}{2} \left[ \frac{p_{a}^{2}}{a} + a - \bar{V} a^{3} \right]\,.
\end{eqnarray}
On quantization, by replacing $p_{a} = - i  \left(\d/\d a\right)$, one gets the Wheeler-de Witt equation for the wave function of the scale factor 
\begin{eqnarray}
\left[ \frac{\d^{2}}{\d a^{2}} + \frac{\gamma}{a} \frac{\d}{\d a} - U(a) \right] \psi(a) = 0\,,
\end{eqnarray}
where $\gamma$ is a constant due to the ambiguity in operator-ordering and
\begin{eqnarray}
U(a) = a^{2} \left( 1 - \bar{V} a^{2} \right).
\end{eqnarray}
In the semi-classical regime, the ambiguity due to operator ordering is not that important and can be ignored in a first approximation \cite{Vilenkin:1987kf}. It is straightforward to see that the behavior of the system in the classically allowed region $U < 0$ (hence, $a > 1/\sqrt{\bar{V}}$) and in the classically forbidden region $U > 0$ (hence, $a < 1/\sqrt{\bar{V}}$) are different. For the classically allowed region, the solution is essentially oscillatory and can be a superposition of in-going and out-going modes; for the classically forbidden region, the solution is a superposition of exponentially growing and decaying modes.

In order to extract a specific solution from these general solutions, one needs to impose boundary conditions. However, quantum cosmology is a \textit{closed} system in which a set of boundary conditions cannot be determined by the environment external to the setup as is the normal practice. In general, there is no fundamental principle to assign all the boundary conditions necessary to specify  the wave function  of the universe $\Psi$. At best, the general consensus is that these boundary conditions need to be supplied as additional fundamental laws of nature. There are two famous, mathematically consistent wavefunctions corresponding to the following boundary conditions:
\begin{enumerate}
	\item \textit{The Hartle-Hawking proposal} \cite{Hartle:1983ai} -- If we choose the \textit{exponentially growing mode} for $a < 1/\sqrt{\bar{V}}$, then the wave function becomes a superposition of in-going and out-going modes for $a > 1/\sqrt{\bar{V}}$.
	\item \textit{The tunneling proposal} \cite{Vilenkin:1984wp} -- If we choose the \textit{out-going mode} for $a > 1/\sqrt{\bar{V}}$, then the wave function becomes a superposition of growing and decaying modes for $a < 1/\sqrt{\bar{V}}$.
\end{enumerate}
The probability of the universe nucleating from `nothing' mainly depends on the contribution from the classically disallowed (hence, quantum) regime. Therefore, the (leading-order contribution to the) probability distribution is approximately 
\begin{eqnarray}
P(a,\phi) \simeq \exp \left({\pm \frac{3}{8V(\phi)}}\right),
\end{eqnarray}
where $+$ corresponds to the Hartle-Hawking wavefunction, and $-$ to the tunneling wavefunction, respectively.

\begin{figure}
\begin{center}
\includegraphics[scale=0.3]{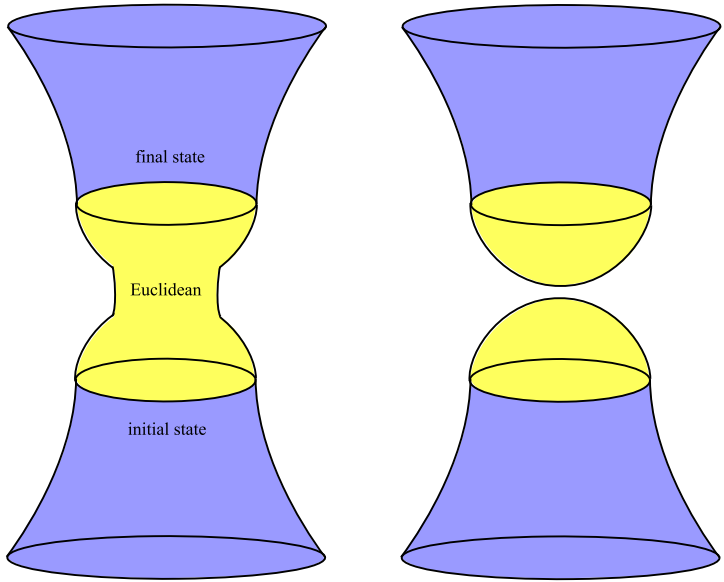}
\caption{\label{fig:E0}Left: Euclidean path integral that connects from the initial state to the final state. Right: If two states are disconnected at the Euclidean manifold, one can consider a wave function for the only final state.}
\end{center}
\end{figure}

\subsection{Euclidean path integral}
Since there is no fundamental principle to choose a boundary condition for the universe in general, we can look towards the path integral quantization for some guidance. In this paper, we shall take this route of quantizing gravity in the path integral formulation instead of the canonical one outlined above. We believe this is the most elegant presentation of the original Hartle-Hawking proposal \cite{Hartle:1983ai} and shall, therefore, stick to it for the LQC-modified case \cite{Brahma:2018elv}. However, as an aside, we note that the Wheeler-de Witt equation mentioned above transforms into a ``difference'' equation (of finite step-size) in LQC resulting from a modified Hamiltonian constraint on a non-separable Hilbert space arising in the theory. It is completely natural that one might impose boundary conditions analogous to the ones mentioned above for this difference equation in a canonical quantization \cite{Bojowald:2003mc}.

Between two hypersurfaces (which take given values of the $3$-metric and scalar field specified on it) at the initial boundary ($h_{\mathrm{i}}$, $\phi_{\mathrm{i}}$) and the final one ($h_{\mathrm{f}}$, $\phi_{f}$), the (Feynmann) propagator is given by
\begin{eqnarray}
\Psi\left[h_{\mathrm{f}}, \phi_{\mathrm{f}}; h_{\mathrm{i}}, \phi_{\mathrm{i}}\right] = \int \mathcal{D}[g] \mathcal{D}[\phi] \;\; e^{i S[g,\phi]}\,,
\end{eqnarray}
where we sum over all geometries, allowing for topology-changes, which have the specified initial and final boundaries. This integration is highly-oscillatory and ill-defined, but its convergence properties can perhaps be improved on introducing a Wick-rotation to Euclidean time $\d t = -i \d\tau$,
\begin{eqnarray}
\Psi_{0}\left[h_{\mathrm{f}}, \phi_{\mathrm{f}}; h_{\mathrm{i}}, \phi_{\mathrm{i}}\right] = \int \mathcal{D}[g] \mathcal{D}[\phi] \;\; e^{- S_{\mathrm{E}}[g,\phi]},
\end{eqnarray}
where now we sum over all Euclidean geometries and corresponding field combinations with the given boundaries (left of Fig.~\ref{fig:E0}). For usual quantum field theories, this corresponds to the ground state wavefunction. Although there is no straightforward way to define the ground state in quantum gravity, but Hartle and Hawking proposed that the above form of the Euclidean path integral may correspond to the ground state wavefunction of the universe.

Let us further assume that the intermediating Euclidean geometries between the initial and final boundaries are disconnected (right of Fig.~\ref{fig:E0}). Especially, if we consider a closed universe (i.e., $k = +1$ FLRW cosmology), then the wavefunction remains well-defined if we remove the initial boundary
\begin{eqnarray}
\Psi_{0}\left[h_{\mathrm{f}}, \phi_{\mathrm{f}}\right] = \int \mathcal{D}[g] \mathcal{D}[\phi] \;\; e^{- S_{\mathrm{E}}[g,\phi]}.
\end{eqnarray}
This unique boundary condition has been given the so-called nickname \textit{no-boundary proposal}, because it has no initial boundary.

There is a lot of justifiable controversy whether this Euclidean path integral is a good approximation of the original Lorentzian path integral or not \cite{Feldbrugge:2017kzv,Feldbrugge:2017fcc,DiazDorronsoro:2017hti,Vilenkin:2018dch}. However, this no-boundary proposal, defined as an Euclidean path integral, is attractive because of several nice properties it possesses.
\begin{itemize}
	\item The Euclidean path integral can be interpreted as the partition function of a thermal system \cite{Gibbons:1976ue}
	\begin{eqnarray}
	Z = \mathrm{Tr} \exp \left({- \beta \hat{\mathcal{H}}}\right) = \int \mathcal{D}[g] \mathcal{D}[\phi] \;\; e^{- S_{\mathrm{E}}[g,\phi]},
	\end{eqnarray}
	where $\hat{\mathcal{H}}$ is the quantum Hamiltonian of the system and $\beta$ is the inverse of the Hawking temperature. The right-hand side can be evaluated in the steepest-descent approximation as
	\begin{eqnarray}
	Z = \exp  \left(-\beta \mathcal{F}\right) \simeq \exp\left( + \frac{3}{8 V_{0}}\right) = e^{\mathcal{A}/4},
	\end{eqnarray}
	where $\mathcal{F} = E - T\mathcal{S}$ is the Helmholtz free energy, $E$ and $\mathcal{S}$ being energy and entropy of the system, respectively. Here, $\mathcal{A}$ is the area of the cosmological horizon. Since the ADM energy $E$ is zero for dS space, we can consistently recover the Bekenstein-Hawking entropy formula $\mathcal{S} = \mathcal{A}/4$. This reveals that the Euclidean path integral is consistent with the expected thermodynamic properties of gravity.
	
	\item Building on this result, one may consider other semi-classical effects in dS space as well. The classical (Lorentzian) equation of motion for a scalar field on a fixed dS background is given by 
	\begin{eqnarray}
	\ddot{\phi} = - 3 H \dot{\phi} - V'\,,
	\end{eqnarray}
	where $H = \dot{a}/a$. On inserting random `white noise' in the slow-roll limit as a thermal effect due to Hawking temperature, one obtains the Langevin equation \cite{LeBellac,Hwang:2012mf}
	\begin{eqnarray}
	\frac{\d}{\d t} \phi = - \frac{V'}{3H} + \frac{H^{3/2}}{2\pi} \xi(t)\,,
	\end{eqnarray}
	where one imposes the white noise conditions:
	\begin{eqnarray}
	\langle \xi(t) \rangle &=& 0\,,\\
	\langle \xi(t) \xi(t') \rangle &=& \delta (t-t')\,.
	\end{eqnarray}
Then, the probability to have the field value $\phi$ at $t$ will follow the Fokker-Planck equation \cite{Linde:1993xx}
	\begin{eqnarray}
	\frac{\partial P(\phi, t)}{\partial t} = \frac{2\sqrt{2}}{3\sqrt{3\pi}} \frac{\partial}{\partial \phi} \left[ V^{3/4}(\phi) \frac{\partial}{\partial \phi} \left( V^{3/4}(\phi) P(\phi,t) \right) + \frac{3V'(\phi)}{8V^{1/2}(\phi)}P(\phi, t) \right]\,,
	\end{eqnarray}
	while the probability to have the field value initially $\chi$ at $t=0$ will follow the equation
	\begin{eqnarray}
	\frac{\partial P(\phi, t|\chi)}{\partial t} = \frac{2\sqrt{2}}{3\sqrt{3\pi}} \left[ V^{3/4}(\chi) \frac{\partial}{\partial \chi} \left( V^{3/4}(\chi) \frac{\partial P(\phi,t|\chi)}{\partial \chi} \right) - \frac{3V'(\chi)}{8V^{1/2}(\chi)}\frac{\partial P(\phi,t|\chi)}{\partial \chi} \right].
	\end{eqnarray}
In the static limit, a solution that satisfies both of these equations is given by
	\begin{eqnarray}
	P(\phi, t|\chi) \sim V^{-3/4}(\phi) \exp \left[ \frac{3}{8V(\phi)} - \frac{3}{8V(\chi)} \right]\,.
	\end{eqnarray}
We can interpret this as the tunneling probability of a homogeneous part of a universe that tunnels from the field value $\chi$ to $\phi$ via stochastic quantum fluctuations when the wavelength is of the order of the Hubble radius.
	
This wave function is consistent with the Euclidean path integral approximated  by the Hawking-Moss instantons \cite{Hawking:1981fz}. On further normalizing the initial boundary, one can obtain the no-boundary wave function. Therefore, we can conclude that the Euclidean path integral describes the stationary limit, or thermal equilibrium, of quantum fluctuations of the Hubble-scale wavelength modes consistently, whereas in many situations, such a thermal equilibrium is coincident with the ground state of the system \cite{Hwang:2012mf}.
\end{itemize}
The above physical motivations are reason enough to investigate and apply the Euclidean path integral with the no-boundary condition as the wavefuntion of the universe, not only due to its mathematical simplicity  but also due to its self-consistency in the low-energy limit with quantum field theory in curved spacetime.

\subsection{Semiclassical approximation: instanton solutions}
One can calculate the Euclidean path integral in the saddle-point approximation by using Euclidean on-shell solutions, or so-called instantons, as
\begin{eqnarray}
\int \mathcal{D}[g] \mathcal{D}[\phi] \;\; e^{- S_{\mathrm{E}}[g,\phi]} \simeq \sum_{\mathrm{instanton}} e^{- S_{\mathrm{E}}^{\mathrm{instanton}}}\,.
\end{eqnarray}
The on-shell solution in pure dS space, on imposing the no-boundary condition $a(0) = 0$, becomes regular to give
\begin{eqnarray}\label{HHInst}
a(\tau) = \frac{1}{H_{0}} \sin \left(H_{0} \tau\right)\,,
\end{eqnarray}
where $H_{0}^{2} = 8\pi \bar{V}/3$. This solution reveals that at the $a(0) = 0$ (``South Pole'') point, we need to impose the condition $\dot{a} = 1$ from the Hamiltonian constraint.

\begin{figure}
\begin{center}
\includegraphics[scale=0.3]{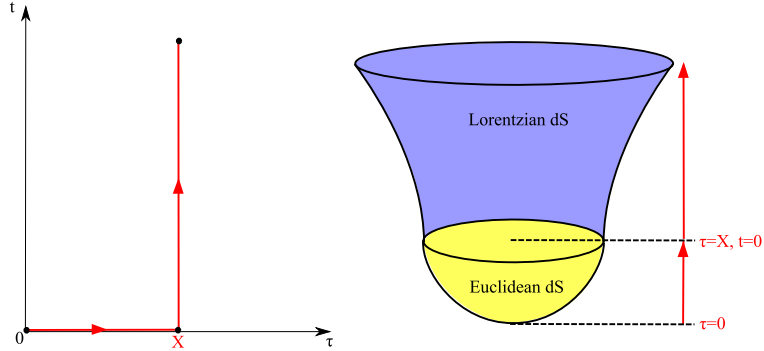}
\caption{\label{fig:E1}Left: A typical time contour over the complex time, where $X = \pi/2H_{0}$. Right: Euclidean and Lorentzian manifold along the given time contour.}
\end{center}
\end{figure}

Inserting this solution, one can evaluate the Euclidean action. In the phase space formulation, the action takes the form
\begin{eqnarray}
S_{\mathrm{E}} = \int \d\tau\; \mathcal{L} = \int \d\tau \left( p_{a} \dot{a} + p_{\phi} \dot{\phi} - N \mathcal{H} \right)\,.
\end{eqnarray}
On plugging the on-shell condition, $\mathcal{H} = 0$ is automatically satisfied. Hence,
\begin{eqnarray}
S^{\mathrm{instanton}}_{\mathrm{E}} = \int \d\tau \left( p_{a} \dot{a} + p_{\phi} \dot{\phi} \right)\,.
\end{eqnarray}
This solution can be analytically continued to Lorentzian time for any constant-$\tau$ hypersurface, but after the Wick-rotation, the metric is in general complex-valued except at the $\dot{a} = 0$ hypersurface (i.e., $\tau = \pi/2 H_{0}$, left of Fig.~\ref{fig:E1}). If we Wick-rotate on this surface, then the metric becomes (right of Fig.~\ref{fig:E1})
\begin{eqnarray}\label{Lorentzian}
a(t) = \frac{1}{H_{0}} \cosh \left(H_{0} t\right)\,.
\end{eqnarray}
We give more mathematical details corresponding to this solution in the following section.

\section{Geometry of the Hartle-Hawking instantons}
We give a more detailed derivation of the schematics described for the minisuperspace model in the previous section. The Friedmann equation (or the Hamiltonian constraint) for the $k=1$ FLRW universe, in Euclidean time, is given by 
\begin{eqnarray}
      \frac{1}{N^2}\left(\frac{\d a}{\d\eta}\right)^2 =\frac{8\pi}{3} \frac{a^2}{N^2}\left(\frac{\d\phi}{\d\eta}\right)^2 - 1 + \left(\frac{\Lambda}{3} + V(\phi)\right) a^2\,,
\end{eqnarray}
where we use the metric $\d s^2 = N^2(\eta)\d\eta^2 + a^2(\eta)\,\d\Omega_3^2$ and set $G=1$ throughout\footnote{Note the new normalization chosen here to facilitate comparison with LQC later on.}. $\eta$ denotes the Euclidean time parameter while reserving $t$ for Lorentzian time, as before. We rewrite the other relevant equation which is the scalar field equation (also, in Euclidean time)
\begin{eqnarray}
\frac{1}{N^2} \left(\frac{\d^2\phi}{\d \eta^2}\right) + \frac{3}{a N}\left(\frac{\d a}{\d\eta}\right) \left(\frac{\d \phi}{\d\eta}\right) - V'(\phi)= 0\,.
\end{eqnarray}
First of all, let us make a gauge choice and fix the lapse function $N=1$. As pointed out in \cite{Hartle:2008ng}, this can be rigorously achieved by introducing the complex variable $\tau(\eta) = \int_0^\eta \d\eta'\, N(\eta')$. Given any lapse function, the variable $\tau$ defines a complex contour on the $\tau$-plane. Once we rewrite the above equations in terms of the variable $\tau$, the task of finding the no-boundary instantons is to solve these equations for the pair of complex analytic functions $a(\tau)$ and $\phi(\tau)$, given the appropriate boundary conditions.

The set of equations, in terms of this new variable, can be expressed as
\begin{eqnarray}
 \dot{a}^2 = 1-\left(\frac{\Lambda}{3} + V(\phi) +\frac{8\pi}{3} \dot{\phi}^2\right) a^2 &=:& \mathcal{V}(a)\label{GREqn1}\,, \\
\ddot{\phi} + 3 H \dot{\phi} -  V'(\phi) &=& 0\label{GREqn2}\,, 
\end{eqnarray}
where a dot refers to a derivative with respect to $\tau$ and the Hubble parameter $H:=\dot{a}/a$. The fact that we can write the RHS of \eqref{GREqn1} as a function of the scale factor alone is only possible for the simplest case of a massless scalar or a cosmological constant. There is always the Raychaudhuri equation involving the second derivative of $a$ but only two of these three equations are linearly-independent. For our purposes of examining the on-shell Euclidean instantons, required for estimating the path integral by its saddle-points, considering these two equations is sufficient. The usual procedure is to solve the above equations for the `no-boundary' boundary conditions \cite{Hartle:2007gi}: $a(0) = 0$ and $\dot{\phi}(0) = 0$. The first condition is a requirement that the geometry must close in a regular fashion while the second is a necessary condition for keeping the solution for $\phi$ regular as $a\rightarrow 0$ \cite{Chen:2016ask}. It is often customary to quote another condition $\dot{a}(0) = 1$; however, this is the consequence of the Friedmann equation. In general, requiring that the scale factor and the scalar field take some fixed value on the final surface, $(a(\tau_f) = b, \phi(\tau_f)=\chi)$, and some fixed value initially, $(a(0) = 0, \dot{\phi}(0) = 0)$, exhausts all the conditions necessary to give a unique solution. The value of the derivative of the scalar field must be fixed from the scalar field equation \eqref{GREqn2} to be zero while the value of the scalar field at the `South Pole' --  $\phi(0) = \phi_0$ -- gives the one-parameter family of instantonic solutions which satisfy the no-boundary proposal \cite{Hartle:2008ng,Hartle:2007gi}. Additional tunings are necessary to ensure the classicality of our universe at late-times, the details of which are unimportant for our purposes (see \cite{Hartle:2008ng,Hwang:2012bd}).

If we take the pure gravity model, in the absence of any scalar field, one can analytically solve for the Euclidean instanton to find that there is a $O(5)-$symmetric solution given by  $a(\tau) = \sqrt{3/\Lambda}\sin\left(\sqrt{\Lambda/3}\,\tau\right)$ (this is Eqn. \eqref{HHInst} above written in the new normalization). In the presence of a scalar field, one requires that the potential is sufficiently flat, i.e. of the inflationary type, for the solution to be regular. In the case of a slowly varying potential, the solution for $a(\tau)$ is a deformed version of the sine function. However, for the massless scalar field (i.e. in the absence of any potential term at all), there are no compact instantons which would give rise to a nontrivial universe. This is obvious from the fact that in this case, the scalar field is in the ``no roll'' condition and the energy density of the universe is trivial. However, this scenario, which is beyond the scope of this paper, leads to new solutions for the no-boundary proposal in the presence of LQC corrections \cite{Brahma:2018elv,Upcoming}.

Before going on to the LQC instantons, let us revisit the geometry of these Hartle-Hawking instantons in Einstein gravity. (This is schematically shown in the right panel of Fig.~\ref{fig:E0}.) Restricting to the case of pure gravity is already sufficient to illustrate its salient features.  For Lorentzian signatures, one has the usual dS solution of Einstein's equations with a positive cosmological constant as $\d s_{\rm L}^2 = -\d t^2 + a_{\rm L}^2(t) \d\Omega^2$, with $a_{\rm L}(t) = \sqrt{3/\Lambda}\cosh\left(\sqrt{\Lambda/3}\;t \right)$ (equivalent to \eqref{Lorentzian} above in the new normalization). To get the $O(5)$ invariant Euclidean instanton from this result, one analytically continues $t$ such that $\tau = \tau_f + it$, where $\tau_f$ is the point where $a(\tau)$ reaches its maximum value. In other words, one gets a Lorentzian dS spacetime from a Euclidean hemisphere (right of Fig.~\ref{fig:E1}), by matching the two hypersurfaces across the zero-extrinsic curvature $(\dot{a}=0)$ `bounce' surface. The effective potential, $\mathcal{V}(a)$, goes to zero on this surface. Such a sharp transition between a real Euclidean half-sphere and a real Lorentzian part is only possible for the simplest example of a pure cosmological constant considered in this paper. In general, in the presence of an inflationary-type potential, the transition would be in terms of `fuzzy' Euclidean instantons \cite{Hartle:2008ng,Hartle:2007gi}, whereby the solutions would be complex in some parts \cite{Hwang:2011mp}. These details, however, are not important for us while focusing near the `South Pole' to exhibit the general `shuttle-cock' type shape of the no-boundary instanton in Euclidean gravity. 

For the pure dS model, the Friedmann equation, in Euclidean time, takes the form $\dot{a}^2 = 1 - \Lambda a^2/3$ which clearly shows that as $a \rightarrow 0$, one gets $\dot{a} \rightarrow 1$. The geometric interpretation of this result goes as follows. The Euclidean $4$-metric, possessing the $O(5)$ symmetry, $\d s^2 = \d\tau^2 + a^2\,\d\Omega_3^2$ has to smoothly close-off in a regular manner into flat space (written in spherical coordinates) $\d s^2 = \d r^2 + r^2\,\d\Omega_3^2$. For this to happen, one has to identify $a(\tau) \sim \tau$ as $a\rightarrow 0$. This suggests that $\dot{a} \rightarrow 1$ in this limit, as is required from the Hamiltonian constraint. However, as mentioned before, this requirement for the derivative of the scale factor automatically follows from the constraint and is not part of the no-boundary condition. 

Let us make one last comment before presenting our new results for the LQC instantons. The no-boundary initial condition is simply that the geometry closes off smoothly, as denoted by $a(0) = 0$. This shall be important later on for the LQC instantons. If we try to solve the Friedmann equation, we get the (famous) unique solution \eqref{HHInst} only on \textit{imposing the no-boundary condition}. Of course, choosing the `initial' point at $\tau =0$ is only for convenience. A priori, there is no need for such a condition to be satisfied by the modified field equations in LQC. However, as we shall show in the LQC case, the initial condition that compact  (Euclidean) instantons in LQC go off to $a \rightarrow 0$ is favored \textit{naturally} even in the presence of quantum-geometry corrections, at least in the pure gravity case. This is the main result of our work which we shall elaborate on in the following sections.

\section{No-boundary instantons in LQC}
In \cite{Brahma:2018elv}, it was shown how the effective LQC action modifies the instantons in the theory even for a simple cosmological constant. Moreover, this leads to slight enhancement of the probability of nucleation of the dS universe from nothing due to the LQC corrections. The general shape of the instanton is reproduced in Fig~\ref{fig:E1}. However, we shall only be interested in the small-$a$ behaviour of this LQC instanton. In particular, what emerges to be intriguing is the infinite tail of the instanton. Although this infinite tail is in Euclidean time, and therefore not physically relevant directly, it does have certain distinguishing features which we shall demonstrate below.

We begin with the modified Friedmann in LQC \cite{Ashtekar:2006es} due to the quantum geometry corrections mentioned in the Introduction.
\begin{eqnarray}\label{ModFriedmann1}
\dot{a}^2 = - a^2 \left(\frac{8\pi}{3}\right) \left(\frac{f^2(a)}{v^2(a)}\right) \left[\tilde{\rho} -  \rho_1\right] \left[\frac{1}{\rho_c}\left(\rho_2 - \tilde{\rho}\right) \right] =: \mathcal{V}_{\rm LQC}(a),
\end{eqnarray}
where $\tilde{\rho}$ is the contribution from a positive cosmological constant. 
\begin{eqnarray}
\tilde{\rho} &:=&\left(\frac{v(a)}{f(a)}\right) \frac{\Lambda}{8\pi}\,,\\
\rho_1 &:=& -\rho_c \left[\sin^2(\sqrt{\Delta}/a) - (1+\gamma^2) \frac{\Delta}{a^2} \right]\, ,\\
\rho_2 &:=& \rho_c \left[\cos^2(\sqrt{\Delta}/a) + (1+\gamma^2) \frac{\Delta}{a^2} \right]\,,
\end{eqnarray}
where
\begin{eqnarray}
v(a) &:=& K \left( \frac{3}{4 \pi \gamma l_{Pl}^2}\right)^{3/2} a^3 V_0\,,\\
f(a) &:=& \left(\frac{1}{2}\right) v(a) ||v(a)-1|- |v(a)+1||\,,\\
K &=& \frac{2\sqrt{2}}{3\sqrt{3\sqrt{3}}}\,,\\
\rho_c &=& \frac{3}{8\pi \gamma^2 \Delta}\,,\\
\Delta &=& 2\sqrt{3} \pi \gamma l_{Pl}^2\,,
\end{eqnarray}
with $V_0 = 16\pi^2$. Although we have adhered to the conventions of \cite{Ashtekar:2006es}, we have generalized their results by adding in non-perturbative expressions for the inverse-triad corrections.

On first look, the above set of equations look rather complicated due to the different terms involved. Here, $f(a)$ represents the inverse-triad corrections whereas holonomy modifications show up in $\rho_1$ and $\rho_2$.  Importantly, note that due to the presence of the  inverse-triad corrections, it is always possible to impose the no-boundary condition $a \rightarrow 0$. However, to gain some intuition into the modified equation, let us begin by setting the holonomy corrections to zero for simplicity. This would be like taking the area gap $\Delta$ to zero. In this limit, $\Delta\rightarrow 0$, we get
\begin{eqnarray}
	\dot{a}^2 = -a^2 \left(\frac{\Lambda}{3}\right) \left(\frac{f(a)}{v(a)}\right) + \left(\frac{f(a)}{v(a)}\right)^2\,.
\end{eqnarray}
This is the modified Friedmann equation only in the presence of inverse-triad corrections. It is easy to check that in the large $a\gg 1$ limit, one gets $f(a) \approx v(a)$, and therefore we get back the usual Friedmann equation for a closed universe. However, in the $a \ll 1$ limit, one cannot make such an approximation. Instead, in this limit, we get $f(a) \approx v(a)^2$. Our aim in this work is not to solve for those instantons which extremizes the Euclidean path integral but rather to examine its small-$a$ behaviour. Therefore, considering $v(a)\propto a^3$ and reinstating the holonomy modifications, we get the leading order term for $a \approx 0$ as
\begin{eqnarray}
	\dot{a}^2 \sim  C a^2\,,
\end{eqnarray}
for some constant $C > 0$. To obtain this result, we notice that the leading order term comes from the $\tilde{\rho}^2$ term in (\ref{ModFriedmann1}) whereas the remaining terms are subdominant. This is a term which arises only in the quantum-corrected Friedmann equation (there is no term quadratic in the energy density in the classical Friedmann equation). The $\tilde{\rho}^2$ term comes with an additional minus sign which leads to a $C>0$. Moreover, note that the dominant contribution in the classical case comes from the curvature term $(1/a^2)$ whereas that term, contained in $\rho_1$, is now sub-dominant. As already argued in the previous section, the essential requirement of the no-boundary condition is the geometry should be closed off in a regular manner and the condition on $\dot{a}$ should follow from the Hamiltonian constraint. In the LQC case, the modified Hamiltonian constraint implies $\dot{a} = 0$ instead of $1$. Nevertheless, $\mathcal{V}(a)$ remains regular even in this case.

\begin{figure}
	\begin{center}
		\includegraphics[scale=0.3]{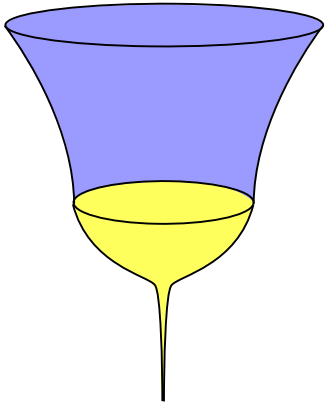}
		\caption{\label{fig:E2}Euclidean and Lorentzian manifold with LQC corrections.}
	\end{center}
\end{figure}

The above findings for no-boundary instantons in LQC in quite remarkable. In order to appreciate this properly, let us make a few comments. Firstly, note that there was no reason that the modified Friedmann equation have to allow for the $a\rightarrow 0$ limit to be imposed consistently. It could easily have been that this limit is singular in LQC. To illustrate this, let us consider only the holonomy modifications while ignoring the inverse-triad ones. Typically, for the Lorentzian effective trajectories such an approximation is completely justified and valid even near the `bounce' surface. In this case, the RHS of \eqref{ModFriedmann1} has a singular term coming from `$\rho_1 \rho_2$' (proportional to $\Delta^2$), which is absent in the classical case. However, luckily for us, when one considers Euclidean histories as is required for our case, one cannot ignore the inverse-triad corrections any longer. Secondly, the structure of the modified Friedmann equation is such that the resulting instantons remain regular for all values of $\tau$ even on imposing the no-boundary condition. For a counterexample, imagine if the form of the equation was such that $\dot{a}^2 \propto a^n$ with $n>2$, in that case, the limit $\tau\rightarrow 0$ would have been singular and there would not have been consistent no-boundary instantons in the theory. Moreover, these inverse-$a$ modifications not only play a crucial role in ensuring that the no-boundary condition can be imposed but also modify the geometry of these instantons to distinguish them from the Einstein gravity. It could also have been the case that the inverse-$a$ modifications are such that one still gets the same condition for $\dot{a}$ at the South Pole. In that case, although the explicit solutions of the instantons would have been different, there would have been no difference in the geometry of LQC and Einstein gravity instantons. The quantum-geometry corrections in LQC \textit{conspire} to ensure that we have no-boundary instantons in the theory with such a geometry that is tapers off to the symmetry point in a novel fashion.

Concretely, the small-$a$ solution for $a$ is given by $a_0 e^{c\tau}$. Obviously, this implies that the point $\tau =0$ is not a good point to impose the no-boundary condition. Rather both $(a(\tau), \dot{a}(\tau))$ goes to zero as $\tau \rightarrow -\infty$. However, this does not represent any difficulty since this infinite stretching is in Euclidean time and is, thus, not physically relevant directly. This novel feature of LQC instantons can be seen from  Fig.~\ref{fig:E2}, where the tail of the compact instanton is stretched infinitely, asymptotically tapering off to zero. The tail \textit{does} contribute to the probability of nucleation of the Lorentzian dS universe, although the path integral and consequently the probability remains finite and well-defined in spite of this infinitely stretched geometry. 

\begin{figure}
\begin{center}
\includegraphics[scale=0.7]{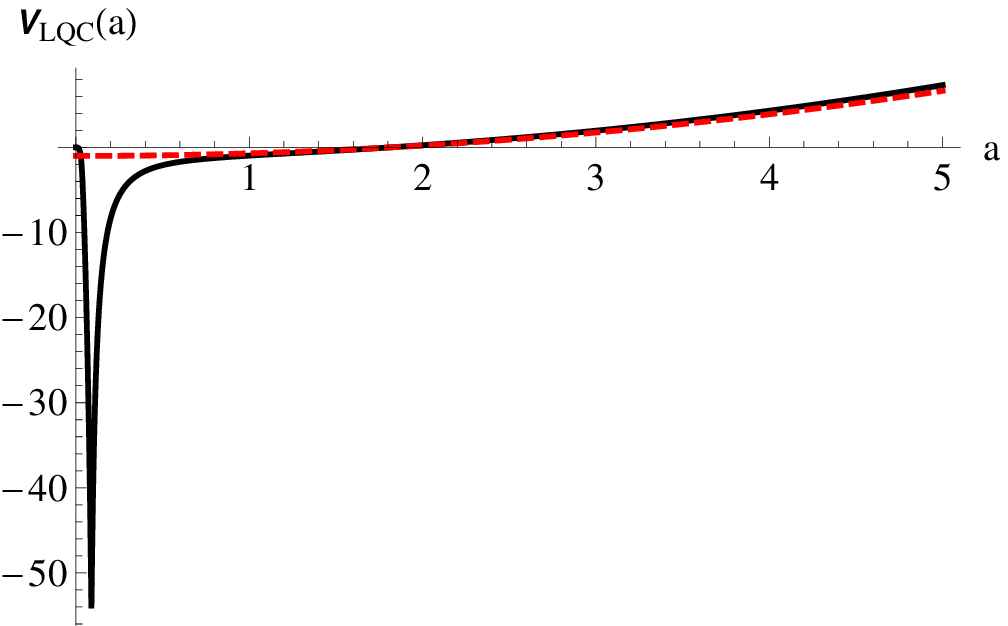}
\includegraphics[scale=0.7]{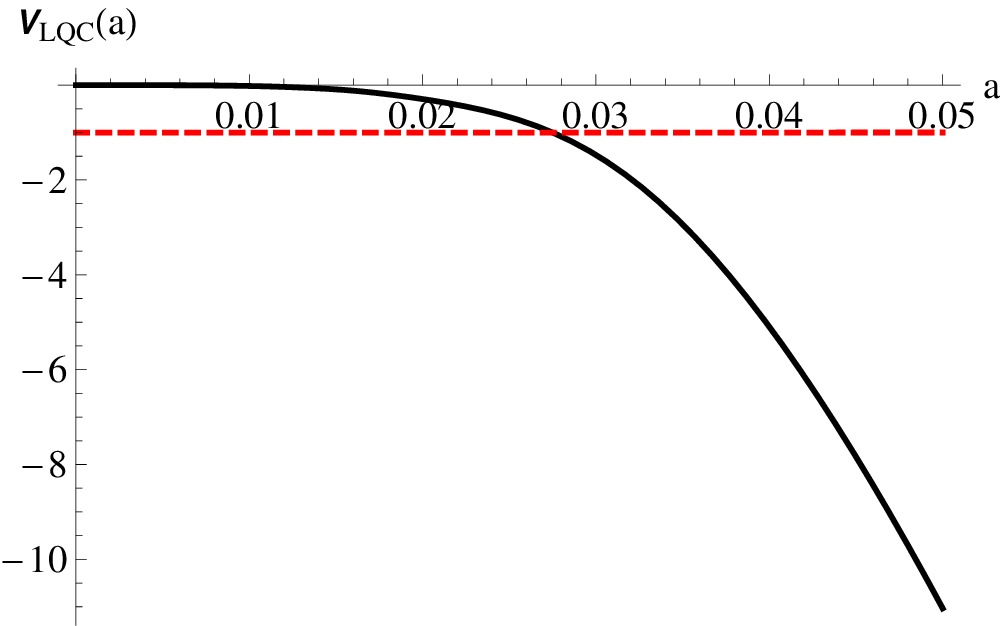}
\caption{\label{fig:pot}The black curve is an example of $\dot{a}^{2}$ for $\Lambda = 1$, $G = 1$, and $l_{Pl} = 0.1$, where the red dashed curve is the limit of the Einstein gravity with the same $a_{\mathrm{max}}$ that satisfies $\dot{a}_{\mathrm{max}} = 0$. Right is the behavior near the $a = 0$ limit.}
\end{center}
\end{figure}

\begin{figure}
\begin{center}
\includegraphics[scale=0.7]{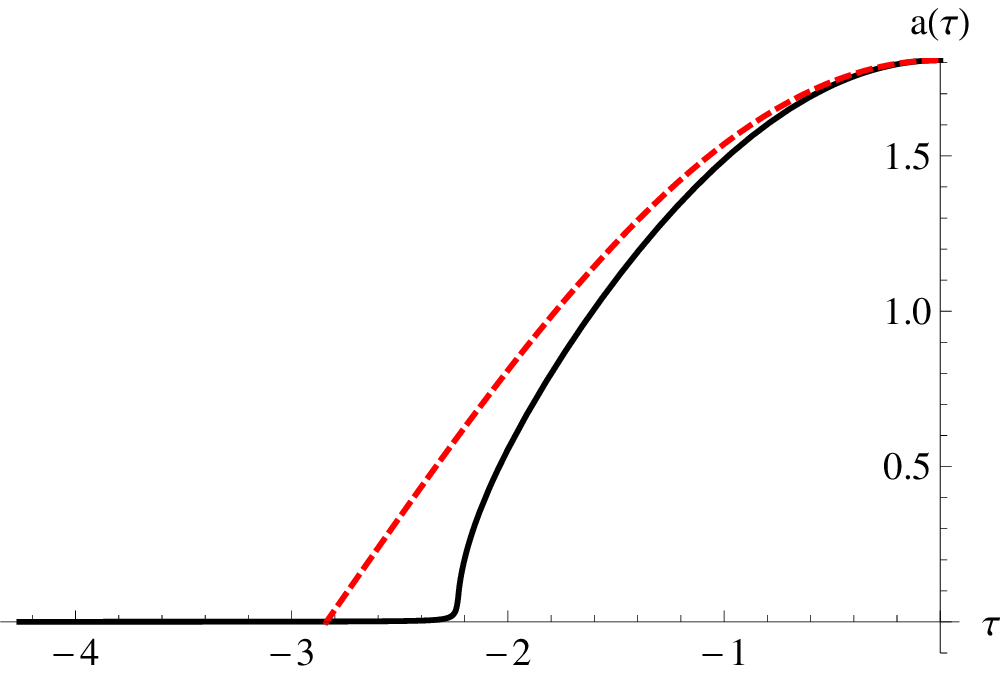}
\includegraphics[scale=0.7]{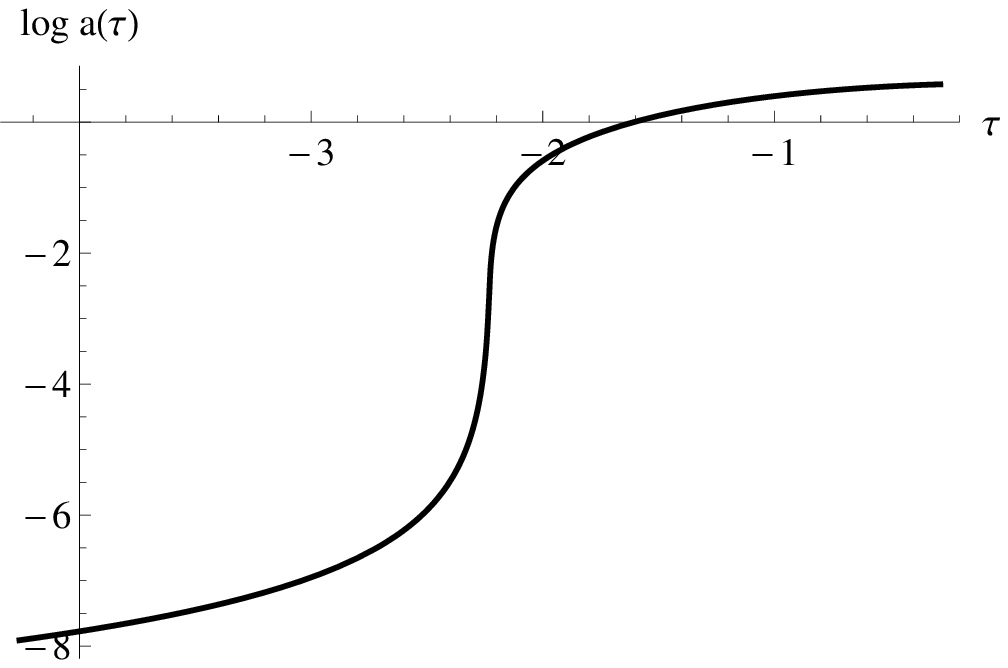}
\caption{\label{fig:inst}Left: $a(\tau)$ (red dashed curve is the limit of the Einstein gravity). Right: $\log a$ for small $a$ limit. It has an infinitely long throat.}
\end{center}
\end{figure}

\subsection{Numerical results}
\begin{figure}
\begin{center}
\includegraphics[scale=0.7]{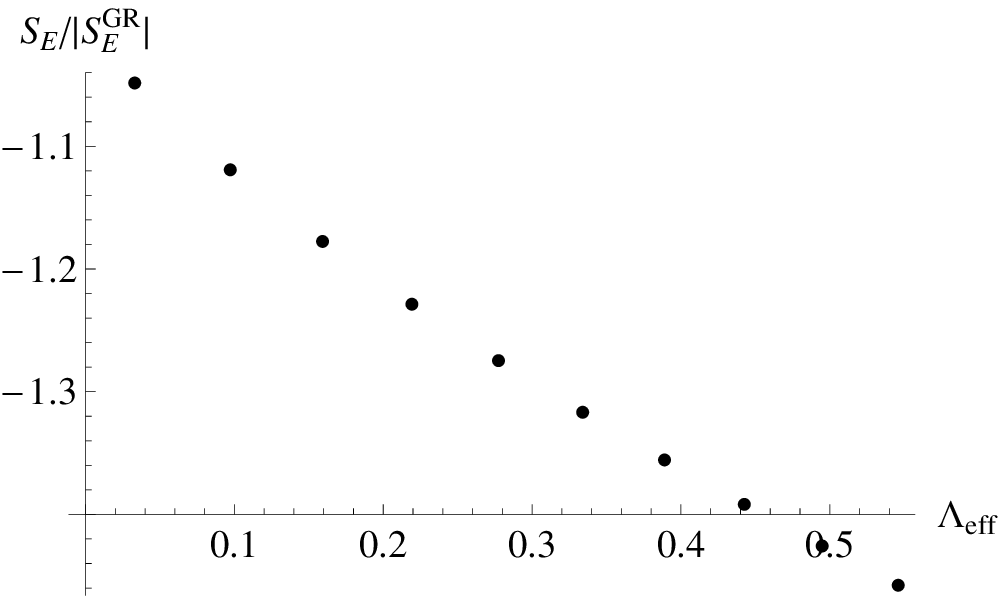}
\includegraphics[scale=0.7]{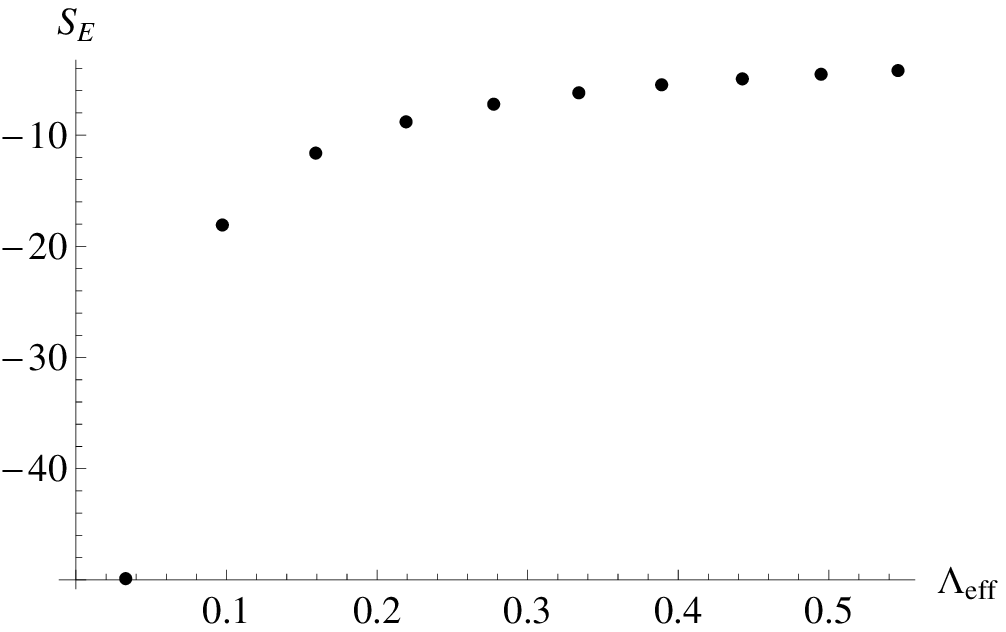}
\caption{\label{fig:action}Left: $S_{\mathrm{E}}/|S_{\mathrm{E}}^{\mathrm{GR}}|$, where for $G = 1$, and $l_{Pl} = 0.1$ by varying $\Lambda$ (equivalently, varying $\Lambda_{\mathrm{eff}} \equiv 1/a_{\mathrm{max}}^{2}$), where $S_{\mathrm{E}}^{\mathrm{GR}}$ is the Euclidean action for the corresponding Einstein limit. Right: $S_{\mathrm{E}}$ for the same parameters.}
\end{center}
\end{figure}

\begin{figure}
\begin{center}
\includegraphics[scale=0.7]{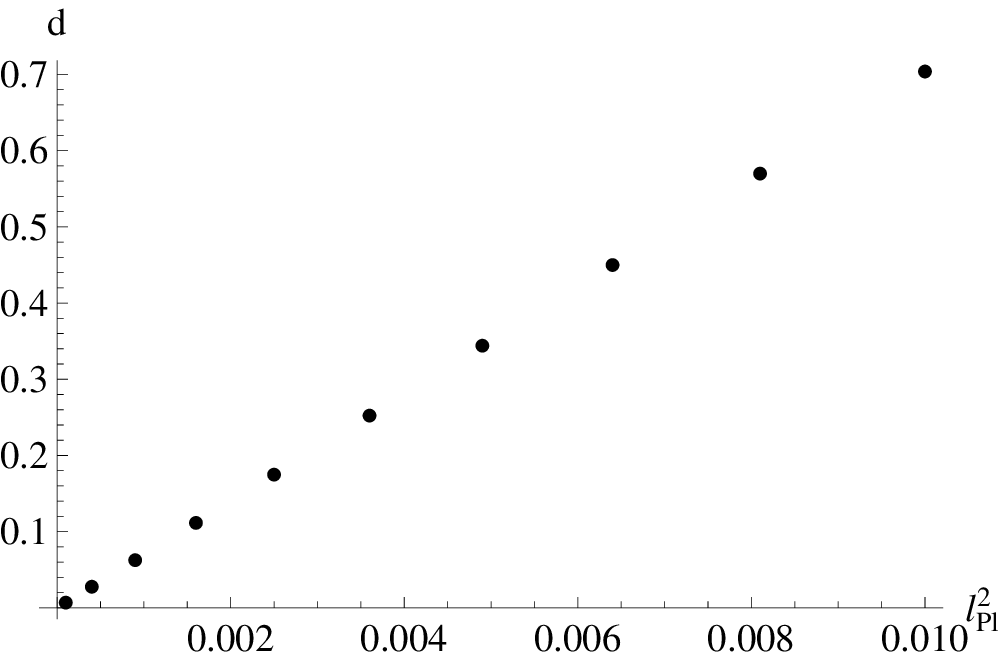}
\includegraphics[scale=0.7]{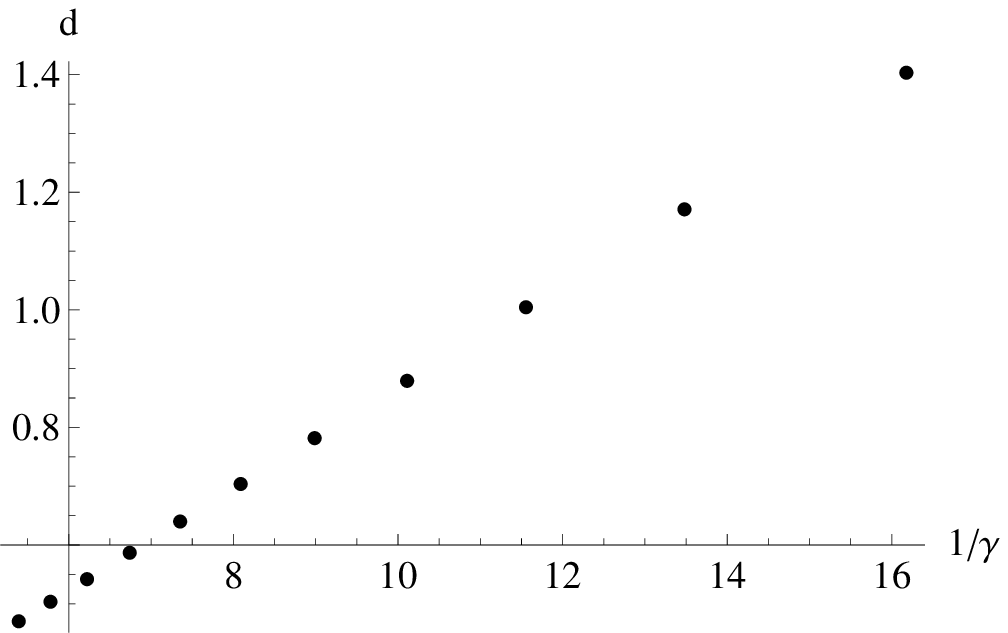}
\caption{\label{fig:action3}Left: By varying $l_{Pl}$, one can calculate $d \propto l_{Pl}^{2}$ numerically. Right: By varying $\gamma$ (with $l_{Pl} = 0.1$), one can see a linear dependence $d \propto 1/\gamma$. We can numerically conclude that $d \simeq 8.7 \times l_{Pl}^{2}/\gamma$.}
\end{center}
\end{figure}

We give some sample numerical solutions for the LQC instantons, going beyond the small-$a$ limit, to illustrate our claims regarding their infinite tail. Figs.~\ref{fig:pot} and \ref{fig:inst} show a typical shape of the effective potential $\mathcal{V}_\text{LQC}(a)$ and its solution $a(\tau)$ for the HR, respectively. This solution demonstrates that the instanton is indeed infinitely stretched (Fig.~\ref{fig:E2}). The approximate behavior of the nucleation probability is of the form (Fig.~\ref{fig:action})
\begin{eqnarray}
S = -2 S_{\mathrm{E}} \simeq \frac{\mathcal{A}}{4} + c + d \log \mathcal{A} + ...
\end{eqnarray}
with a model dependent positive constant $c$, where $\mathcal{A} = 4 \pi a_{\mathrm{max}}^{2}$. As mentioned before, $a_\text{max}$ denotes the max value the instanton takes in the Euclidean regime, on which surface we analytically continue to the Lorentzian regime. The above constant $c$ can easily be absorbed away in the normalization of the probability measure  and the only relevant correction due to LQC, over the Einstein-Hilbert value, comes from the parameter $d$. This parameter $d$ can be approximately expanded by $d\simeq 8.7 \times l^{2}_{Pl} / \gamma$ (Fig.~\ref{fig:action3}) in terms of the fundamental parameters -- Planck length and Immirzi parameter -- of the theory, as shown via numerical reconstruction.


\section{\textit{Robustness} of the no-boundary condition}
In Einstein gravity, when considering a positive cosmological constant as the only matter source, the only solution which is regular at the South Pole is given by \eqref{HHInst}. The crucial point here is that one gets this solution for the instanton in Einstein gravity from the (Euclidean) Friedmann equation on imposing the no-boundary condition. Interestingly, as was shown in previous section, the small-$a$ behaviour for the LQC instantons following the modified Friedmann equation is always of the form
\begin{eqnarray}
	a(\tau) \propto e^{C \tau}\,,  
\end{eqnarray}
for some constant $C$. This shows that as $\tau \rightarrow -\infty$, we get $a\rightarrow 0$ implying a natural implementation of the no-boundary condition in LQC instantons, as a result of quantum-geometry regularizations, at least for the simplest case of a cosmological constant. In this sense, this shows that the no-boundary condition is robust and more natural when the Einstein-Hilbert action is augmented by LQG corrections.

The Euclidean instantons, in this pure gravity scenario, are always going to be compact and there is no risk of an Euclidean wormhole forming for some values of the parameter space. This conclusion is true both for Einstein gravity as well as the LQC case. Therefore, one always gets compact Euclidean instantons for the pure dS case in both cases. The overall conceptual picture of the nucleation of the universe from nothing is also the same in both cases. One quantitative difference is due to the modified equations in LQC: The `bounce' surface ($\dot{a}=0$), as predicted by the effective trajectories of sharply-peaked semiclassical states, in LQC is (slightly) different from the hypersurface joining the Euclidean and Lorentzian parts in Einstein gravity. However, the qualitative behaviour remains the same and this difference is reflected in enhancing the probability of nucleation of the universe manifesting as a next-to-leading-order correction in LQC. As mentioned in \cite{Brahma:2018elv}, these type of terms can also appear in Einstein gravity from going to higher order corrections and there should be a competition between the two terms in LQC -- one appearing from radiative corrections and the other from the inherent quantum geometry. It is thus difficult to unambiguously state that LQC enhances the tunneling amplitude for such no-boundary universes.

\section{Conclusion}
It is an old expectation that \textit{mathematical consistency} alone shall be sufficient to derive the boundary conditions in quantum cosmology \cite{Bojowald:2001xa}. The form of the LQC instantons suggest that it might indeed be possible to identify \textit{typical} smooth initial conditions due to quantum-geometry corrections. This demonstrates that one of the  most fundamental proposal for the initial condition in quantum cosmology can appear naturally in another  -- LQC -- due to putative corrections coming from quantum geometry. The introduction of the no-boundary proposal in LQC has also opened new physical possibilities for the latter. Instead of replacing the big bang singularity with a deterministic bounce, as is predicted by semiclassical states in some restricted models of LQC, this opens up the opportunity to allow for Euclidean trajectories leading to a bubble nucleation of our universe. The semiclassical saddle-point approximation of the no-boundary proposal is distinct from the semiclassical sharply-peaked states in LQC and therefore the former acts as an example of how a new state can unveil novel features in a well-established theory. At this point, it is difficult to compare the probability of a bounce versus that of tunneling of the universe from nothing. However, the no-boundary proposal also provides a (set of) \textit{natural} initial conditions for considering inhomogeneous perturbations in LQC leading to effects observable from early-universe cosmology without having to resort to \textit{ad hoc} choices for the initial state. 

Regarding the geometry of no-boundary instantons in LQC, we have demonstrated that the feature of having an infinite tail distinguishes these instantons from the original Hartle-Hawking ones in Einstein gravity. We end our discussion with a few caveats. Firstly, it has been pointed out recently that the Euclidean path integral in gravity is not a good approximation for the original Lorentzian path integral due to several conceptual reasons \cite{Feldbrugge:2017kzv}. However, even if one works with the Lorentzian path integral and applies a different mathematical trick (Pecard-Lefshetz theory) to improve its convergence, the resulting theory typically has runaway perturbations due to the old conformal factor problem in gravity \cite{Feldbrugge:2017fcc}. Interestingly, LQC can come to the rescue of the no-boundary proposal \cite{Upcoming2}, written as a Lorentzian path integral, even in this case. However, the main physical effect from LQC responsible for this is `dynamical signature-change' \cite{Bojowald:2015gra}, something we have ignored in this work as a first pass. The second caveat is regarding the fact that our discussions were limited to the case of a pure cosmological constant in this paper. Indeed, the more interesting physical scenario is that of having a scalar field in some potential. However, preliminary investigations have already revealed that the solution space of the no-boundary wavefunction is greatly enhanced for such a system in the presence of LQG corrections. Finally, one can ask how physical is the fact that the tail of these instantons are stretched to infinity? As already mentioned, this is only true in Euclidean time and therefore not directly meaningful. However, it might even be possible that for some different gauge choice (i.e. $N\neq 1$), one can even avoid such an infinite stretching altogether. Nevertheless, all the interesting effects of having such a geometry as explained in this paper would still be valid in this case. Most importantly, no matter what the gauge choice, the remarkable conclusion that the modified Friedmann equation in LQC not only allows for the no-boundary condition to be imposed but also somehow makes it more natural seems to be robust and points towards a new paradigm in quantum cosmology merging these two mainstream approaches.

\section*{Acknowledgements}
This research was supported in part by the Ministry of Science, ICT \& Future Planning, Gyeongsangbuk-do and 
Pohang City and the National Research Foundation of Korea grant no. 2018R1D1A1B07049126.

\end{document}